\documentclass[11pt,twoside]{article}
\usepackage{asp2010}
\usepackage{graphicx}

\resetcounters

\begin{document}

\title{Polar structures in late-type galaxies}
\author{Alexei Moiseev
\affil{Special Astrophysical Observatory, \fbox{Russian Academy of Sciences}, Nizhnij Arkhyz, Russia}
}
\begin{abstract}
A common point of view is that  stable polar structures are mainly present
in  the  E/S0 early-type galaxies  lacking dense gas in their host
disks. However,  nuclear, as well as external
polar rings and disks also exist in the  late-type hosts, including
gas-rich dwarf galaxies.  Using the 3D spectroscopic observations of
these objects we can derive the rotation properties of the gas
components separately in the main and polar planes.  The most detailed
picture of the gas kinematics and mass distribution properties could
be obtained from the combination of optical (H~{\sc ii}) and radio (H~{\sc i}) data
sets. I  briefly review the results of such studies, including the observations of the direct interaction  between the multi-spin gas components.
\end{abstract}

\section{Introduction}

When talking about a polar ring galaxy (PRG), we usually imply  a system, where a central early morphological type red galaxy is surrounded by a ring, similar  to the late-type objects, having blue colors  and  containing gas and young stellar population:   \textit{``..in all PRGs the H~{\sc i} gas is associated with the polar structure and not with the central stellar spheroid...''} \citep{2006ApJ...643..200I}. Indeed,  in the case  of early-type central objects the matter accreted  from the environment with an orthogonal orientation of the orbital moment has never experienced any direct collisions with the pre-existing ISM. However late-type hosts also exist among PRGs \citep[][ and references below]{1995AJ....109..942V}. The SDSS-based catalog of PRG candidates \citep{2011MNRAS.418..244M} lists the galaxies, where both components with different spins have relative blue colors, i.e. posses H~{\sc ii} regions and young stellar population: SPRC~52, SPRC~171, SPRC~269, etc. These  examples show that the interaction  between two gaseous  subsystems does not dramatically destroy the global structure of a large-scale polar disk, at least in some objects.  In contrast with  external polar rings where the most of the central galaxies are E/S0, the inner polar structures (IPS)  nested  in a wide range of morphological types:  about $30\%$ of confirmed  IPS are observed in the  Sb--Im galaxies \citep{2012AstBu..67..147M}. This implies a larger real fraction of late-type PRGs, as the inner and large-scale polar structures form a single family in the distribution of normalized diameters, but the methods of candidate  selection vary (image inspection for PRGs, study of circumnuclear kinematics for IPS).

For a long time  PRGs are considered  to be  a  good probe to study the dark matter 3D-shape \citep*{1987ApJ...314..439W,2003ApJ...585..730I}. In the case of ``classical'' PRGs,  special attempts are needed to reconstruct the  circular rotation curve from the early-type galaxy stellar kinematics: deep absorption-line spectroscopy and  subsequent asymmetric-drift correction which  depends on the accepted dynamical model parameters  and quality of  velocity dispersion measurements. In contrast with early-types hosts,  rotation curves of both components in the late-type PRG could  be  reconstructed from the  emission-line velocity field derived from  the radio interferometric observations of neutral and molecular gas, or optical observations in the ionized gas  emission lines with scanning Fabry-Perot interferometers (FPI) or other 3D-spectroscopy devices.

\begin{figure}
\centerline{
\includegraphics[height=0.55\textwidth]{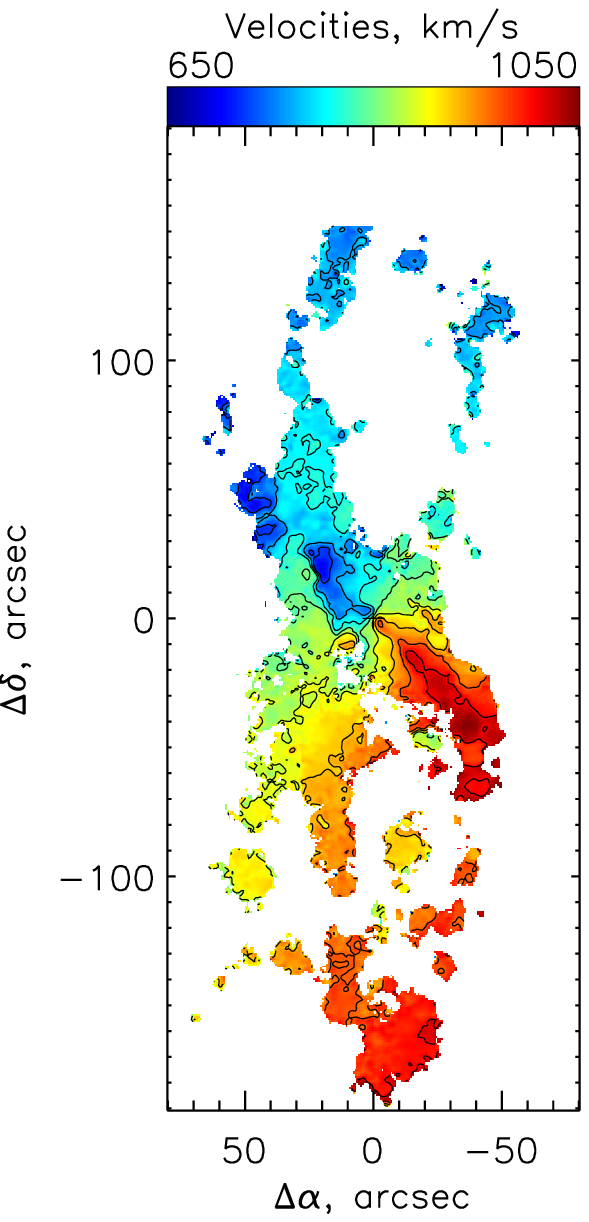}
\includegraphics[height=0.55\textwidth]{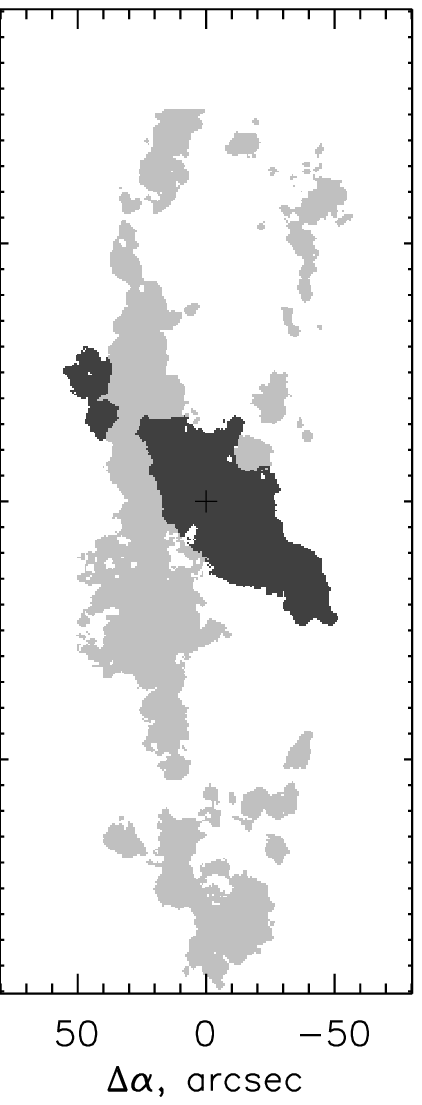}
\includegraphics[height=0.55\textwidth]{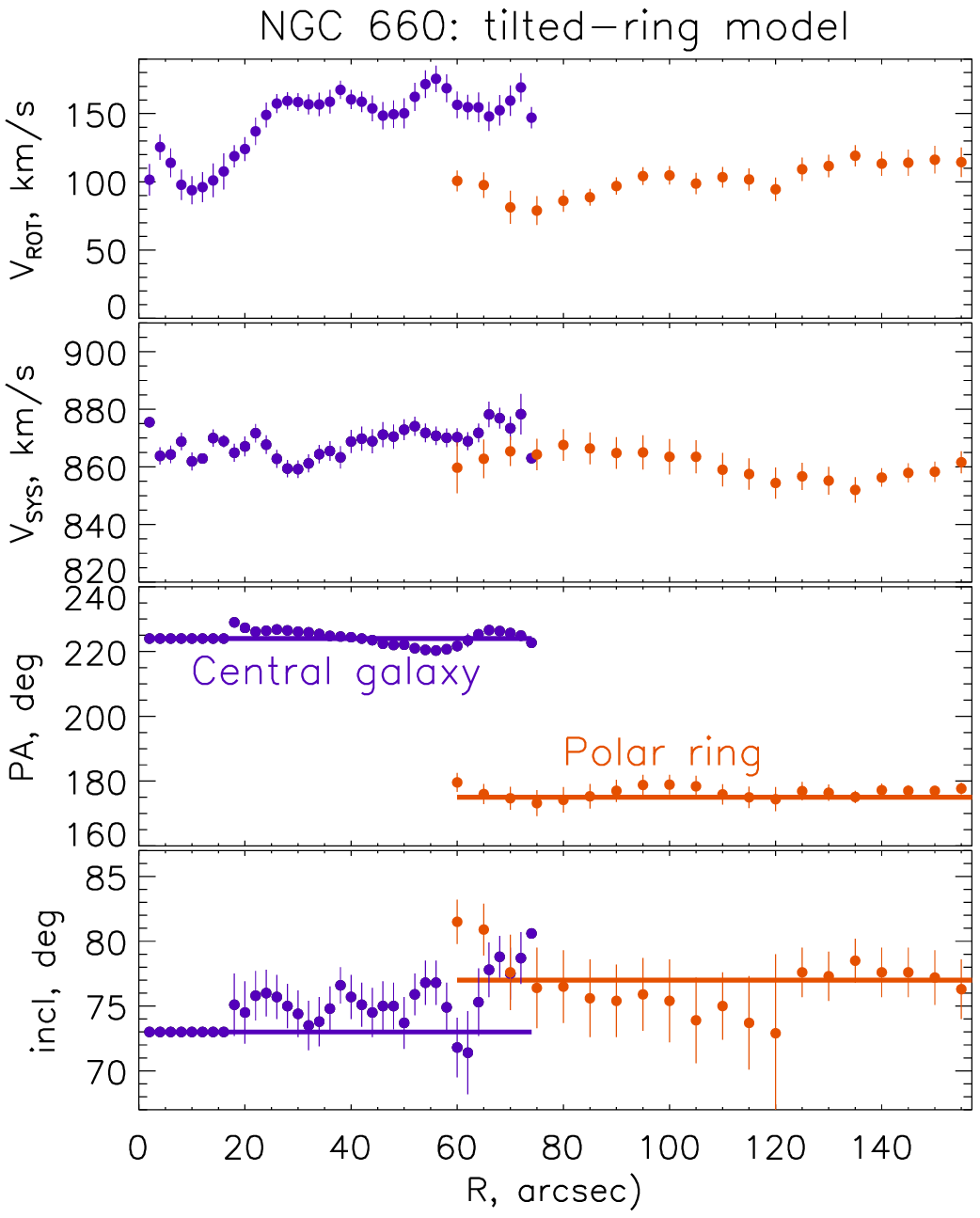}
}
\caption{Ionized gas kinematics in NGC 660: left -- the H$\alpha$ velocity field obtained  at the SAO RAS 6-m telescope with the scanning FPI; middle -- the subdivision of this field into the domains of the inner disk (dark gray) and polar ring  (light shaded). The right plot shows the results of tilted-ring analysis for  both kinematical components: radial variation of the rotation velocity,   systemic velocity,   position angles and  inclination. The solid lines mark the accepted values of $PA$ and $i$.
}
\label{001Moiseev_fig1}
\end{figure}

\section{Two gaseous rotating components}

Fig.~\ref{001Moiseev_fig1} shows the velocity field of ionized gas in  NGC 660  derived from the observation at the SAO RAS 6-m telescope with the scanning FPI in the SCORPIO focal reducer \citep{AfanasievMoiseev2005}. Using the tiled-rings approximation  we were able   to study separately the kinematics of the central disk and the external  structure, which is inclined by the angle of about $50\deg$ to the main disk. The ring of ionized gas is flat ($PA,i\approx const$), while a warp is observed in the H~{\sc i} for the larger radii \citep{1995AJ....109..942V}. The maximal  rotation velocity of the central disk is about $1.5$ times larger than that of the ring. Such behaviour   contradicts with the common trend of PRGs where the rings rotate faster than the hosts, as it follows from  their position on the Tully-Fisher diagram  \citep*{2003ApJ...585..730I,2013A&A...554A..11C}. Now we are matching  the ionized gas kinematics   with the new WSRT H~{\sc i} data to estimate of the dark matter shape in NGC 660.

The interesting examples  of  gas rotation in both planes are the post-merger systems like UGC 5600 \citep{2007AstL...33..520S}, Arp~230 and MCG~-5-7-1 \citep{2013AJ....145...34S}. Based on the properties of H~{\sc i} kinematics in two latter galaxies, the authors claimed that the  dark halo potential is spherical, with only   $5-10\%$ flattening.

\section{Streams interaction and gas ionization state}

The analysis of NCG~660 velocity field demonstrates an overlapping of the central disk and the ring at the distances  $r\approx$60--$75''$ (4--5 kpc). It means that the orbits of matter with different spins  have intersected. Moreover, gaseous clouds can avoid meetings at these cross ways during many revolutions, because only   a few H{\sc II} regions are located on the overlapping radii. Also, the width of the ``collision zone'' depends on the orbits   ellipticity. The more intriguing case  is the Arp 212  galaxy which provides the evidence of direct collisions between the ISM in the main galaxy plane and in the internal part of the inclined  polar ring: a peculiar dust lane is observed in the region where the orbits  (calculated  from FPI velocity field analysis) intersect each other \citep{2008AstBu..63..201M}. The integral-field observations of this galaxy  also point to a possible contribution of shocks into the gas ionization, as it follows from the emission-line ratio maps [N~{\sc ii}]/H$\alpha$ \citep{2008ApJ...677..201G} and  [O~{\sc iii}]/H$\beta$ \citep{2012A&A...547A..24C} for the region of the matter collision.  In this case the shock fronts related with the interaction between multi-spin gaseous components provoke the formation of dense molecular clouds and trigger the burst of star formation observed in Arp~212 now.

The shock waves should be formed in  the classical PRGs too. \citet{1993AJ....105.1745W} suggested that the gas on the polar orbits should experience shocks passing trough the gravitational well of the stellar   disk, even in the case at the  central disk lacks  gas. Unfortunately,  direct observation of shocks in  PRGs is still an open question, while the shock contribution in the gas ionization is crucial for  the gas metallicity estimation using emission lines ratios.

Some circumstantial evidences of shock waves formation among PRGs  are provided by the recent studies of tilted gaseous disks accreted from the galactic environment. The spectroscopic observation  of NGC~7743 have discovered a  disk of ionized gas inclined by $33\deg$ or $77\deg$ to the stellar one \citep*{2011ApJ...740...83K} with the domination of shock waves in the gas  emission lines excitation. The similar effect is also  found in non-coplanar gaseous disks of isolated lenticular galaxies \citep*{2013arXiv1312.6701K}. An  alternative mechanism providing LINER-like emission is the ionization  by post-AGB stars \citep{2010MNRAS.402.2187S}. New spectral observations including detection of   faint emission lines are crucial for the final conclusion.

\section{Polar structures in  dwarf galaxies}

A traditional way of new PRG search  consists in collecting the candidates with the corresponding  optical morphology \citep{1990AJ....100.1489W,2011MNRAS.418..244M} and the follow-up observations of their kinematics. Dwarf star forming galaxies (dIrr, blue compact dwarf galaxies, H~{\sc ii}-galaxies) are a more difficult case, because the irregular distribution of bright knots  is usually a dominating feature in the optical/NIR images. It's not surprising that a discovery of polar structures  in many  dwarf galaxies is a byproduct of study   of the H~{\sc ii} or H~{\sc i} velocity fields: SDSS J102819.24+623502.6 \citep{2009ApJ...696L...6S}, Mrk 370 \citep{2011EAS....48..115M}, DDO~99 and  UGC~8508 \citep{2014AstBu..69M}. A serious problem  is that the velocity pattern in the expanding  bubbles around the sites of star formation can mimic the kinematic behaviour of multi-spin structures \citep*[see the Appendix in][]{2010MNRAS.405.2453M}. Additional information (velocity dispersion distribution, UV/optical/NIR morphology) is needed for a  final  conclusion.

It is belived that interactions are the main mechanism  which provokes bursts of star formation in many dwarf galaxies. It means that a large fraction of  polar structures in these galaxies is related with the same interaction event (accretion or merging). For instance, at least 18\%  of  nearby luminous blue compact dwarf galaxies  in the sample of 28  objects possess external and inner polar structures \citep[see the references in][]{2011EAS....48..115M}.   Surveys of  kinematics of galaxies with  large field of view IFU will discover new examples of such systems.

\acknowledgements The work was supported by the RFBR grant 13-02-00416 and  the ``Active Processes in Galactic and Extragalactic Objects'' basic research program of the Department Physical Sciences of the RAS OFN-17. I am also grateful to the `Dynasty' Foundation.   The observations at the 6-m telescope were carried out with the financial support of the Ministry of Education and Science of Russian Federation (contracts no. 16.518.11.7073 and 14.518.11.7070).


\begin{thebibliography}{}


\bibitem[{{Afanasiev} \& {Moiseev}(2005)}]{AfanasievMoiseev2005}
{Afanasiev}, V.~L., \& {Moiseev}, A.~V. 2005, Astronomy Letters, 31, 194

\bibitem[{{Cair{\'o}s} et~al.(2012){Cair{\'o}s}, {Caon}, {Garc{\'{\i}}a
  Lorenzo}, {Kelz}, {Roth}, {Papaderos}, \& {Streicher}}]{2012A&A...547A..24C}
{Cair{\'o}s}, L.~M., {Caon}, N., {Garc{\'{\i}}a Lorenzo}, B. et al. 2012, \aap, 547, A24

\bibitem[{{Combes} et~al.(2013){Combes}, {Moiseev}, \&
  {Reshetnikov}}]{2013A&A...554A..11C}
{Combes}, F., {Moiseev}, A., \& {Reshetnikov}, V. 2013, \aap, 554, A11

\bibitem[{{Garc{\'{\i}}a-Lorenzo} et~al.(2008){Garc{\'{\i}}a-Lorenzo},
  {Cair{\'o}s}, {Caon}, {Monreal-Ibero}, \& {Kehrig}}]{2008ApJ...677..201G}
{Garc{\'{\i}}a-Lorenzo}, B., {Cair{\'o}s}, L.~M., {Caon}, N., {Monreal-Ibero},
  A., \& {Kehrig}, C. 2008, \apj, 677, 201

\bibitem[{{Iodice} et~al.(2003){Iodice} et~al.}]{2003ApJ...585..730I}
{Iodice}, E., {Arnaboldi}, M., {Bournaud}, F.  et~al. 2003, \apj, 585, 730

\bibitem[{{Iodice} et~al.(2006){Iodice}, {Arnaboldi}, {Saglia}, {Sparke},
  {Gerhard}, {Gallagher}, {Combes}, {Bournaud}, {Capaccioli}, \&
  {Freeman}}]{2006ApJ...643..200I}
{Iodice}, E., {Arnaboldi}, M., {Saglia}, R.~P.  et~al. 2006, \apj, 643, 200

\bibitem[{{Katkov} et~al.(2011){Katkov}, {Moiseev}, \&
  {Sil'chenko}}]{2011ApJ...740...83K}
{Katkov}, I.~Y., {Moiseev}, A.~V., \& {Sil'chenko}, O.~K. 2011, \apj, 740, 83

\bibitem[{{Katkov} et~al.(2013){Katkov}, {Sil'chenko}, \&
  {Afanasiev}}]{2013arXiv1312.6701K}
{Katkov}, I., {Sil'chenko}, O., \& {Afanasiev}, V. 2013, MNRAS, accepted.
  \url{1312.6701}

\bibitem[{{Moiseev}(2011)}]{2011EAS....48..115M}
{Moiseev}, A. 2011, in EAS Publications Series, edited by M.~{Koleva},
  P.~{Prugniel}, \& I.~{Vauglin}, vol.~48 of EAS Publications Series, 115

\bibitem[{{Moiseev}(2008)}]{2008AstBu..63..201M}
{Moiseev}, A.~V. 2008, Astrophysical Bulletin, 63, 201

\bibitem[{{Moiseev}(2012)}]{2012AstBu..67..147M}
{Moiseev}, A.~V. 2012, Astrophysical Bulletin, 67, 147

\bibitem[{{Moiseev}(2014)}]{2014AstBu..69M}
{Moiseev}, A.~V. 2014, Astrophysical Bulletin, 69, in press

\bibitem[{{Moiseev} et~al.(2010){Moiseev}, {Pustilnik}, \&
  {Kniazev}}]{2010MNRAS.405.2453M}
{Moiseev}, A.~V., {Pustilnik}, S.~A., \& {Kniazev}, A.~Y. 2010, \mnras, 405,
  2453

\bibitem[{{Moiseev} et~al.(2011){Moiseev}, {Smirnova}, {Smirnova}, \&
  {Reshetnikov}}]{2011MNRAS.418..244M}
{Moiseev}, A.~V., {Smirnova}, K.~I., {Smirnova}, A.~A., \& {Reshetnikov}, V.~P.
  2011, MNRAS, 418, 244

\bibitem[{{Sarzi} et~al.(2010){Sarzi} et~al.}]{2010MNRAS.402.2187S}
{Sarzi}, M., {Shields}, J.~C., {Schawinski}, K.  et~al.  2010, \mnras, 402, 2187

\bibitem[{{Schiminovich} et~al.(2013){Schiminovich}, {van Gorkom}, \& {van der
  Hulst}}]{2013AJ....145...34S}
{Schiminovich}, D., {van Gorkom}, J.~H., \& {van der Hulst}, J.~M. 2013, \aj,
  145, 34

\bibitem[{{Shalyapina} et~al.(2007){Shalyapina}, {Merkulova}, {Yakovleva}, \&
  {Volkov}}]{2007AstL...33..520S}
{Shalyapina}, L.~V., {Merkulova}, O.~A., {Yakovleva}, V.~A., \& {Volkov}, E.~V.
  2007, Astronomy Letters, 33, 520

\bibitem[{{Stanonik} et~al.(2009){Stanonik}, {Platen}, {Arag{\'o}n-Calvo}, {van
  Gorkom}, {van de Weygaert}, {van der Hulst}, \&
  {Peebles}}]{2009ApJ...696L...6S}
{Stanonik}, K., {Platen}, E., {Arag{\'o}n-Calvo}, M.~A. et~al. 2009,
  \apjl, 696, L6

\bibitem[{{van Driel} et~al.(1995){van Driel}, {Combes}, {Casoli}, {Gerin},
  {Nakai}, {Miyaji}, {Hamabe}, {Sofue}, {Ichikawa}, {Yoshida}, {Kobayashi},
  {Geng}, {Minezaki}, {Arimoto}, {Kodama}, {Goudfrooij}, {Mulder}, {Wakamatsu},
  \& {Yanagisawa}}]{1995AJ....109..942V}
{van Driel}, W., {Combes}, F., {Casoli}, F. et~al. 1995,
  \aj, 109, 942

\bibitem[{{Wakamatsu}(1993)}]{1993AJ....105.1745W}
{Wakamatsu}, K.-I. 1993, \aj, 105, 1745

\bibitem[{{Whitmore} et~al.(1990){Whitmore}, {Lucas}, {McElroy},
  {Steiman-Cameron}, {Sackett}, \& {Olling}}]{1990AJ....100.1489W}
{Whitmore}, B.~C., {Lucas}, R.~A., {McElroy}, D.~B., {Steiman-Cameron}, T.~Y.,
  {Sackett}, P.~D., \& {Olling}, R.~P. 1990, \aj, 100, 1489

\bibitem[{{Whitmore} et~al.(1987){Whitmore}, {McElroy}, \&
  {Schweizer}}]{1987ApJ...314..439W}
{Whitmore}, B.~C., {McElroy}, D.~B., \& {Schweizer}, F. 1987, \apj, 314, 439

\end{thebibliography}

\end{document}